\documentstyle[12pt]{article}

\font\blackboard=msbm10 at 12pt
\font\blackboards=msbm7
\font\blackboardss=msbm5
\newfam\black
\textfont\black=\blackboard
\scriptfont\black=\blackboards
\scriptscriptfont\black=\blackboardss

\def\identity{{\rlap{1} \hskip 1.6pt \hbox{1}}}

\def\Tr{{\rm Tr\ }}
\def\half{\frac{1}{2}}

\newcommand{\NP}{{\em Nucl.\ Phys.\ }}

\newcommand{\PRL}{{\em Phys.\ Rev.\ Lett.\ }}

\newcommand{\tr}{{\rm Tr}}

\newcommand{\gone}[1]{}
\begin{document}
\pagestyle{plain}
\setcounter{page}{1}

\baselineskip16pt

\begin{titlepage}

\begin{flushright}
PUPT-1749\\
hep-th/9712105
\end{flushright}
\vspace{20 mm}

\begin{center}
{\Large \bf 
Longitudinal 5-branes\\
as 4-spheres
in Matrix theory}

\vspace{3mm}

\end{center}

\vspace{10 mm}

\begin{center}
{Judith Castelino, Sangmin Lee and Washington Taylor IV}

\vspace{2mm}

{\small \sl Department of Physics} \\
{\small \sl Joseph Henry Laboratories} \\
{\small \sl Princeton University} \\
{\small \sl Princeton, New Jersey 08544, U.S.A.} \\
{\small \tt judithc, sangmin, wati @princeton.edu}

\end{center}

\vspace{1cm}

\begin{abstract}
We present a construction in Matrix theory of longitudinal 5-branes
whose geometry in transverse space corresponds to a 4-sphere.  We
describe these branes through an explicit construction in terms of $N
\times N$ matrices for a particular infinite series of values of $N$.
The matrices used in the construction have a number of properties
which can be interpreted in terms of the 4-sphere geometry, in analogy
with similar properties of the $SU(2)$ generators used in the
construction of a spherical membrane.  The physical properties of
these systems correspond with those expected from M-theory; in
particular, these objects have an energy and a leading long-distance
interaction with gravitons which agrees with 11D supergravity at leading
order in $N$.
\end{abstract}

\vspace{1cm}
\begin{flushleft}
December 1997
\end{flushleft}
\end{titlepage}
\newpage

\section{Introduction}

One of the most convincing pieces of evidence for the Matrix theory
conjecture \cite{BFSS} is the fact that supersymmetric matrix quantum
mechanics seems to contain most of the extended objects of 11D
supergravity, and that the interactions between these objects are
described by a potential whose leading long-distance behavior agrees
with supergravity.  By now, we have a very good understanding
of how the supermembrane arises in Matrix theory.  This connection was
first understood many years ago by de Wit, Hoppe and Nicolai
who quantized the supermembrane theory in light-front gauge
and arrived at precisely the Matrix theory Lagrangian
\cite{Goldstone-Hoppe,dhn}.  Banks,
Fischler, Shenker and Susskind used this connection as evidence in
their original presentation of the Matrix theory conjecture, where
they constructed infinite membranes in terms of Matrix theory
variables.  It was shown that the infinite membranes of Matrix theory
have the correct long-distance interactions in
\cite{Aharony-Berkooz,Lifschytz-Mathur}.  More recently, it was proven
that any Matrix theory solution corresponding to a classical membrane
configuration interacts with other Matrix theory objects through a
time-averaged potential which depends only on the energy of the state,
amounting to a proof of the equivalence principle in matrix theory
\cite{Dan-Wati}.

Our understanding of the M-theory 5-brane in Matrix theory, however,
is currently much less complete.  There are two ways in which we might
expect a 5-brane to appear in Matrix theory.  Depending upon whether
the 5-brane has been wrapped around the longitudinal direction or not,
the 5-brane will appear in Matrix theory as either a longitudinal
5-brane (L5-brane) which is extended in only four of the nine
transverse Matrix theory dimensions, or as a transverse 5-brane which
is extended in five of the transverse dimensions.  The L5-brane was
described as a Matrix theory background in \cite{Berkooz-Douglas}.  It
was argued in \cite{grt} that an L5-brane can be explicitly described
in Matrix theory variables as an object with a set of four fields
$X^i, i \in \{1, 2, 3, 4\}$ having a nonzero value of the
antisymmetric product
\begin{equation}
{\rm Tr}\;\epsilon_{ijkl} X^i X^j X^k X^l.
\label{eq:4-charge}
\end{equation}
This result was derived using T-duality on a compact 4-torus, and is
simply the dual of the statement that an instanton on a 4-brane
carries 0-brane charge \cite{Witten-small,Douglas}.  This
identification of the L5-brane charge was found independently by
Banks, Seiberg and Shenker \cite{bss}, who calculated the Matrix
theory supersymmetry algebra explicitly and found a central charge
corresponding to (\ref{eq:4-charge}) playing the role of L5-brane
charge.  The fact that infinite L5-branes have the correct
long-distance gravitational interaction was verified in
\cite{Lifschytz-46}.  Thus, we have some understanding of how flat
longitudinal 5-branes appear in Matrix theory.  However, to date there
has been little progress in understanding how to describe L5-branes
which are embedded in any other way than as flat 4-surfaces in
transverse space.

The goal of this paper is to construct L5-brane configurations which
are not flat, but which correspond to configurations in transverse
space with the topology of a 4-sphere.  We have succeeded in doing
this only for certain values of $N$.  Our solutions correspond to
completely symmetric spherical configurations.  Although these
solutions seem to have all the properties we would expect of spherical
(really, cylindrical) M-theory 5-branes, we have not found a
convenient way of including fluctuations on the surfaces of these
L5-branes which break the spherical symmetry.  This is rather
different from the situation for the membrane 2-sphere.  In the case
of the membrane, there are particularly simple solutions in the
spherically symmetric case which are described using the
$N$-dimensional $SU(2)$ generators in $U(N)$ \cite{Dan-Wati,Rey}.
However, using the formalism of de Wit, Hoppe and Nicolai it is
possible to include fluctuations by writing the matrices describing
the spherical configuration as symmetrized polynomials in the $SU(2)$
generators.

The structure of this paper is as follows: In Section 2 we present the
details of the Matrix 4-sphere construction.  This construction is
highly analogous to the construction of a symmetric membrane 2-sphere
in Matrix theory in terms of $SU(2)$ generators.  In Section 3 we
discuss the properties of the Matrix 4-sphere solution, including the
equations of motion, energy, 4-brane charge and gravitational
interactions.  We find that all calculations agree with expectations
from M-theory.  We also discuss some difficulties in trying to
extend this construction to geometries with arbitrary fluctuations.
Section 4 contains conclusions.

Note added: It was pointed out to us by Klim\v{c}\'{i}k that the
L5-brane 4-sphere construction described here is closely related to a
``fuzzy'' 4-sphere which has been considered in the context of 4D
field theory on spaces with noncommutative geometry \cite{gkp}.

\section{Matrix 4-sphere construction}

\subsection{Properties}

To describe a classical Matrix configuration with the geometrical
properties expected of a 4-sphere, we expect to have five $N \times N$
matrices $X_1, \ldots, X_5$ which are nonzero and have certain
properties which can be interpreted in terms of the 4-sphere geometry.
These properties are analogous to similar properties discussed in
\cite{Dan-Wati} for the membrane 2-sphere, where the Matrix membrane
2-sphere is described by the $SU(2)$ generators $J_i$ as in
\cite{dhn}.

Before giving the explicit Matrix realization of the 4-sphere, let us
list the properties we will expect of the matrices $X_i$.  For each
property we discuss the analogous property satisfied by the
matrices $Y_i = 2rJ_i/N$ describing the spherical Matrix membrane.

\vspace{0.2in}
\noindent
{\bf  I. Spherical locus} 

We expect that the 0-branes composing the membrane should be
constrained in a noncommutative sense to lie on a 4-sphere of radius
$r$.  This corresponds to the algebraic condition
\[
X_1^2 + X_2^2 + X_3^2 + X_4^2 + X_5^2 = r^2 \identity.
\]
This condition is analogous to the condition
\[
Y_1^2 + Y_2^2 + Y_3^2 = r^2 \identity  +{\cal O} (1/N^2)
\]
satisfied by the matrices describing the spherical
Matrix membrane.  We only expect this relation to hold to leading
order in $1/N$ due to the fuzzy nature of geometry at finite $N$.

\vspace{0.1in}
\noindent
{\bf II. Longitudinal 5-brane charge}

It was shown in \cite{grt,bss} that
an L5-brane in Matrix theory is locally described by a set
of 4 matrices $Z_1, \ldots, Z_4$ satisfying
$\epsilon^{ijkl}Z_iZ_jZ_kZ_l = \alpha \identity$ where $\alpha$ is a
constant.  So that the matrices $X_i$ have the correct L5-brane
charge, we impose the condition
\begin{equation}
\epsilon^{ijklm}  X_i X_j X_k X_l = \alpha X_m
\label{eq:algebra}
\end{equation}
where $m$ is any fixed value between 1 and 5, and where $\alpha$
is a constant to be determined.  This is analogous to the condition
$[Y_i, Y_j] \sim i\epsilon_{ijk}Y_k$ satisfied by the $SU(2)$
generators.

\vspace{0.1in}
\noindent
{\bf II'. Local flatness} 

As a direct consequence of the property (\ref{eq:algebra}) we find
that the geometry of the spherical membrane appears locally to be that
of a flat L5-brane.  For example, by looking at the matrices $X_i$ in
the block where $X_5$ has eigenvalues close to $r$ we find the
relation
\[
\epsilon^{ijkl}  X_i X_j X_k X_l = \alpha r \identity +{\cal O} (1/N)
\]
where the epsilon contraction is over indices 1-4.
This is analogous to the result that the Matrix membrane is locally
described by, for example, $[Y_1, Y_2] \sim i\identity +{\cal O}
(1/N)$ in the block where $Y_3$ has eigenvalues close to $r$.

\vspace{0.1in}
\noindent
{\bf III. Rotational invariance} 

We would like the spherical L5-brane to have a Matrix description
invariant under the $SO(5)$ symmetry of the 4-sphere.  This
corresponds mathematically to the condition that
\[
R_{ij} \cdot X_j = U (R) \cdot X_i \cdot U(R^{-1})
\]
where $R_{ij} \in SO(5)$ is an element of the rotation group and $U
(R)$ is an $N$-dimensional unitary representation of $SO(5)$ (or its
covering group).  This corresponds to the result for the spherical
membrane that the analogous condition holds where $R$ is an element of
$SO(3)$ and $U$ is in a representation of the covering group $SU(2)$.

\vspace{0.1in}
\noindent
{\bf IV. Spectrum} 

We can always diagonalize a single matrix $X_i$.  The
eigenvalues of this matrix correspond to positions of the individual
0-branes on the $i$-axis.
We expect that the 0-branes composing the L5-brane
should have a density in each direction corresponding to the projected
density of a 4-sphere along a single coordinate axis.  This density
should be proportional to $r^2 -x^2$ where $x$ is the position along
the $i$-axis.  For the Matrix membrane this corresponds to the
observation that the matrices $Y_i$ have a spectrum which is uniformly
distributed on the interval $[-r, r]$ as expected for the projected
density of a 2-sphere.

\subsection{Construction}

We now give an explicit construction of a family of matrices $X_i$
which satisfy the above conditions.  We have found such matrices only
for values of $N$ of the form
\[
N = \frac{(n + 1) (n + 2) (n + 3)}{6}.
\]
In the next subsection we discuss the possibility of finding L5-brane
spheres for other values of $N$.

To begin with, consider the $4 \times 4$ Euclidean gamma matrices
$\Gamma_i$, $i \in\{1, \ldots, 5\}$.  An explicit construction of
these matrices in $2 \times 2$ block-diagonal form is given by 
\begin{eqnarray}
\Gamma_i & = &  \left(\begin{array}{cc}
0 & -i \sigma_j\\
i \sigma_j& 0
\end{array}\right), \;\;\;\;\; i \in\{1, 2, 3\} \nonumber\\
\Gamma_4 & = &  \left(\begin{array}{cc}
0 & \identity_2 \\
\identity_2& 0
\end{array}\right) \label{eq:gamma}\\
\Gamma_5 & = &  \left(\begin{array}{cc}
\identity_2 & 0\\
0&  -\identity_2
\end{array}\right) \nonumber
\end{eqnarray}
where $\sigma_j$ are the usual Pauli matrices.
These matrices satisfy the conditions
\begin{equation}
\Gamma_i^2 = \identity,
\label{eq:g1}
\end{equation}
\[
\Gamma_i \Gamma_j = -\Gamma_j \Gamma_i\;\;\;\;\; {\rm for} \;i \neq j,
\]
and
\begin{equation}
\epsilon^{ijklm} \Gamma_i \Gamma_j \Gamma_k \Gamma_l = 24 \Gamma_m.
\label{eq:g3}
\end{equation}
{}From (\ref{eq:g1}) and (\ref{eq:g3}) we see that these matrices
already satisfy conditions I and II which we expect of matrices
describing an L5-brane sphere.  

The commutators of the gamma matrices
\[
\Gamma_{ij} =[\Gamma_i, \Gamma_j]/2 = \Gamma_i \Gamma_j \;\;\;\;\; i
\neq j
\]
are a set of ten $4 \times 4$ matrices which satisfy the Lie algebra of
$SO(5)$.  These matrices form the 4-dimensional spin representation of
this algebra.  Using these matrices we can verify that property III is
satisfied by the matrices $\Gamma_i$.  It remains to be shown that we
can use the $\Gamma$ matrices as building blocks to describe an
infinite family of matrices which still satisfy conditions I, II and III
and whose spectrum of eigenvalues asymptotically approaches $r^2 -x^2$.

To construct our desired matrices $X_i$ which describe the spherical
L5-brane, we simply construct the $n$-fold symmetric tensor product
representation of the $\Gamma$ matrices
\[
G_i^{(n)} =  \left(
\Gamma_i \otimes \identity \otimes \cdots \otimes \identity +
\identity \otimes \Gamma_i \otimes \identity \otimes \cdots \otimes \identity +
\cdots +
\identity \otimes \cdots \otimes\identity \otimes \Gamma_i \right)_{\rm Sym}
\]
where by Sym we mean the restriction to the completely symmetrized
tensor product space.  The dimension of the $n$-fold symmetric tensor
product space is given by
\[
N =\frac{(n + 1) (n + 2) (n + 3)}{6}
\]
and the matrices $G_i^{(n)}$ are therefore matrices of this size.  
A number of useful identities for the $G$ matrices are collected in
Appendix A.
It is clear that the spectrum of eigenvalues for the matrices
$G_i^{(n)}$ contains the set of integers in the range $[-n, n]$.
Thus, we define the matrices $X_i$ describing the 4-sphere through
\begin{equation}
X_i = \frac{r}{n}  G_i^{(n)}
\label{eq:sphere}
\end{equation}
for an arbitrary value of $n$.
We now
proceed to verify that these matrices satisfy each of the desired
properties  I-IV.

Property I follows directly from the matrix identity (\ref{eq:Schur})
\[
\sum_{i}(G^{(n)}_i)^2 = c \identity_N = n(n+4) \identity_N.
\]
Clearly then, 
\[
\sum_{i} X_i^2= \frac{r^2}{n^2} \sum_{i}  (G_i^{(n)})^2
= r^2 \identity_N + {\cal O} (1/n),
\]
as needed for property I.

Without loss of generality, we can verify property II by just checking the
case $m=5$.  Notice first that
\[
\epsilon_{ijkl}X^i X^j X^k X^l = \{[X_1,X_2], [X_3, X_4]\}
+\{[X_2,X_3], [X_1, X_4]\} + \{[X_3,X_1], [X_2, X_4]\},
\]
where indices run from 1 to 4.  Among the $24n^2$ terms on the
right-hand side, the $24n$ terms of the form $\Gamma_{12} \Gamma_{34}
\otimes \identity \cdots \otimes \identity$ combine to make $24
(r/n)^3 X_5$.  Let us illustrate the effect of the $24n(n-1)$
cross-terms in the $n=2$ case.  For $n=2$, we can write the
cross-terms in the following form:
\[
4 \left(\frac{r}{n}\right)^4 \epsilon_{ijk}  
\left\{\Gamma_{ij}\otimes\Gamma_{k4} + \Gamma_{k4}\otimes\Gamma_{ij}\right\}
= 8 \left(\frac{r}{n}\right)^4  
(\identity_2 \otimes \sigma_i) \otimes (\sigma_3 \otimes \sigma_i) 
+ 8 \left(\frac{r}{n}\right)^4  
(\sigma_3  \otimes \sigma_i) \otimes (\identity_2 \otimes \sigma_i), 
\]
where the indices run from 1 to 3. In the last line, we used our
explicit choices for the gamma matrices (\ref{eq:gamma}).
When restricted to the totally symmetric subspace, $\sigma_i\otimes\sigma_i$ 
is easily shown to be equal to $(\identity_2\otimes\identity_2)$. 
Since $X_5$ in this notation is
\[
X_5 = {r\over n}
\left\{ (\sigma_3  \otimes \identity_2) \otimes  
(\identity_2 \otimes \identity_2)
+ (\identity_2  \otimes \identity_2) \otimes (\sigma_3 \otimes
\identity_2) \right\}, 
\]
we see that the cross terms contribute 8 more factors of $ (r/n)^3
X_5$.  In a similar way, one can show for any $n$ that the $24n(n-1)$
cross terms make $8(n-1) (r/n)^3 X_5$. Altogether, we have
\[
\epsilon^{ijklm}  X_i X_j X_k X_l = (8n + 16)\left( \frac{r}{n} \right)^3 X_m.
\]
We see then that (\ref{eq:algebra}) is satisfied for any $n$ with the 
coefficient $\alpha = (8n + 16) (r/n)^3$. The dependence on $n$
of the coefficient may seem strange. We will see later that this 
construction has the charge of $n$ overlapping branes. 

To verify property III, we note that the matrices
\[
G_{ij}^{(n)} =[G_i^{(n)}, G_j^{(n)}]/2
\]
are precisely the generators of the $N$-dimensional representation of
$SO(5)$.  These matrices can be simply exponentiated to get $U (R)$
for any rotation in $SO(5)$.

To verify property IV we need to consider the asymptotic form of the
eigenvalue density of the matrices $X_i$ as $n \rightarrow
\infty$.  We can compute the exact eigenvalue density at finite $n$ by
constructing a canonical basis for the symmetrized product space and
calculating the eigenvalues of $X_5$ on each basis vector.  As a
basis, we can take the vectors
\begin{equation}
| i_1 i_2 \cdots i_n \rangle = {\rm Sym} (u_{i_1} \otimes
u_{i_2} \otimes \cdots \otimes u_{i_n})
\label{eq:eigenvectors}
\end{equation}
where $1 \leq i_1 \leq i_2 \leq \cdots \leq i_n \leq 4$
and where $u_1, \ldots, u_4$ are an eigenbasis of $\Gamma_5$ with
eigenvalues $e_1 = e_2 = 1$, $e_3 = e_4 = -1$.  Each
vector of the form (\ref{eq:eigenvectors}) is an eigenvector of
$G_5^{(n)}$ with
\[ 
G_5^{(n)} | i_1 i_2 \cdots i_n \rangle=
(e_{i_1} + e_{i_2} + \cdots + e_{i_n}) | i_1 i_2 \cdots i_n \rangle\ .
\]
To count the number of eigenvectors with a fixed eigenvalue $m$ we see
that the $e_{i_k}$ eigenvalues must consist of $\frac{n + m}{2}$ +1's and
$\frac{n -m}{2}$ -1's.  There are precisely
\[
\left(\frac{n + m}{2}  + 1 \right) \cdot \left(
\frac{n-m}{2}  + 1 \right) = \frac{(n + 2)^2 -m^2}{4} 
\]
ways of doing this, so this is the number of eigenvectors with
eigenvalue $m$.  Translating this result into the spectrum of $X_5$ we
see that for the $n$th representation the spectrum of $X_5$ is
\[
x_5 = r \frac{m}{n}  \;\;\;\;\; {\rm with\ degeneracy}  \;\;\;\;\;
\frac{(n + 2)^2 -m^2}{4}.
\]
This gives a 0-brane density proportional to $r^2 -x^2$ as $n
\rightarrow \infty$, verifying that the matrices we have constructed
satisfy property IV.

\subsection{Other values of $N$}

We have explicitly constructed spherical Matrix L5-branes only for a
restricted class of values of $N$.  It would be interesting to know
whether it is possible to construct matrices satisfying properties
I-IV for any other values of $N$.  From property III it is clear that
$N$ must be a dimension for which a (spin) representation of $SO(5)$
is possible.  It might seem like the construction given here
generalizes naturally by including some antisymmetrization on the
tensor product space.  However, it turns out that, except for the
totally symmetric cases, the $G_i$'s mix different representations of
$SO(5)$. For example, in the decomposition of $SO(5)$ representations
$\bf 4\otimes 4 = 10 \oplus 5 \oplus 1 \rm$, $G_i$'s transform the
states of the $5-$dimensional representation into that of the
$1-$dimensional representation and vice versa.

As the simplest case of a dimension which corresponds to an $SO(5)$
representation but which does not admit an L5-brane sphere in the form
we have found, consider $N = 5$. The infinitesimal form of the
rotational invariance condition reads
\begin{equation}
\delta^{ik} X^j - \delta^{jk} X^i = [J^{ij}, X^k],
\label{infinitesimal}
\end{equation}
where $J^{ij}$ are the (anti-hermitian) generators of $SO(5)$ whose 
components are given by
\[
(J^{ij})_{lm} = \delta^i_l \delta^j_m - \delta^i_m \delta^j_l.
\]
Using this, we can write down the components of (\ref{infinitesimal}),
\[
\delta^{ik} X^j_{lm} - \delta^{jk} X^i_{lm}
= \delta^i_l X^k_{jm} - \delta^j_l X^k_{im}
+ X^k_{li} \delta^j_m + X^k_{lj} \delta^i_m.
\]
We will see that the only solution to this equation is the trivial one, 
namely, $X^k_{lm} = 0$ for all $k,l,m$. First, put $i=1$, $j=2$, $k=5$, 
$l=1$, $m = 3,4,5$. This gives $X^5_{2m} = 0$ for $m=3,4,5$. 
By choosing similar sets of indices, we can show that all the 
off-diagonal components of $X^k$ are zero.

Each $X^k$ is now a traceless diagonal $5\times5$ matrix. As such they cannot
be linearly independent. So we can find a nonzero 5-vector $u_k$ such that
$u_kX^k = 0$. If we normalize $u_k$ to be a unit vector, $u_kX^k$ is nothing
but the component of the vector $X$ in the $u$-direction. Since all the 
$X^k$'s are related to $u_kX^k$ by some unitary rotation, they should also 
vanish.
This proves that there are no $5\times 5$ matrices $X^i$ satisfying
conditions I-IV.

\section{Physical properties}

\subsection{Energy}

In units\footnote{See Appendix B for our normalization conventions.}
where $l_{11}=1$
the potential energy term in the Matrix Hamiltonian reads
\[
H_{pot} = - R \frac{1}{4 (2\pi)^2} \tr [ X_i, X_j ]^2.
\]
Using the identities in Appendix A, we find that
\[
H_{pot} = \frac{R}{4(2\pi)^2} \frac{r^4}{n^4} 16Nc 
= n T_{M5} (2\pi R) \Omega_4 r^4 + {\cal O}(1/n),
\]
where $T_{M5} = (2\pi)^{-5}$ is the tension of an M5-brane and 
$\Omega_4 = 8\pi^2 /3 $ is the area of a unit 4-sphere.
This is the right value of the energy for  $n$ spherical L5-branes.

\subsection{4-brane charge}

We can check to make sure the configuration corresponds to  $n$
spherical L5-branes by calculating the 4-brane charge of the ``upper
hemisphere'' corresponding to the 0-branes with positive eigenvalues
of $X_5$.

An analogous calculation for the membrane sphere gives,
\[
C_2 = \frac{(2 \pi)^2}{2 \pi} {\rm Tr}_{1/2}\;[Y_1,Y_2] 
=  \pi r^2 +{\cal O} (1/N)
\]
which is at leading order
precisely the area of the projected sphere.

For the 4-sphere we have
\[
C_4 =  \frac{(2 \pi)^4}{8\pi^2} {\rm \Tr}_{1/2} \; \epsilon^{ijkl}
X_i X_j X_k X_l 
= n
 \frac{\pi^2 r^4}{2} + {\cal O} (1/n),
\]
which is $n$ times the area of the projected 4-sphere.

There is another way of calculating the total charge of the brane.
Recall that if we look at the small block of the $X_i$'s where $X_5$
has the largest eigenvalue, the brane is locally flat and the charge
is well-defined. The charge of this block should be identified with
the total charge of the brane times the ratio of the size of the block
to the size of the whole matrix.  For the membrane sphere, this
definition gives
\[
Q_2 = \frac{(2 \pi)^2}{2\pi} \left({2r\over N}\right)^2 {N-1\over 2} \times N 
= 4\pi r^2 + {\cal O}(1/N) ,
\]
which is the area of a 2-sphere. For the 4-sphere,
\[ 
Q_4 =  \frac{(2 \pi)^4}{8\pi^2}(8n +16)\left({r\over n}\right)^4 n\times N 
= n  \frac{8\pi^2 r^4}{3} + {\cal O}(1/n),
\]
which is $n$ times the area of the 4-sphere.

Note that this charge multiplied by $2 \pi R T_{M5}$ is exactly the
same as the energy computed above at leading order in $n$. This
reflects the fact that the brane is locally flat and BPS near any
point on the sphere.  The value of the charge and the energy lead us
to conclude that our construction actually describes $n$ overlapping
spherical longitudinal 5-branes.  As we will discuss in Section
\ref{sec:fluctuations}, it is not clear whether it is possible to
separate the branes from each other.

\subsection{Absence of 2-brane charge}

The local BPS condition derived above implies that the 4-sphere
is a pure L5-brane without 2-brane charge. It is indeed possible
to show that the 2-brane charge vanishes in each block of the matrix
$X_5$ for any $n$.

Each state of the $n$-fold totally symmetric tensor representation can
be labeled by its weight in the $SO(5)$ Cartan subalgebra $G_{12},
G_{34}$,
\[
|V\rangle = {\rm Sym} 
( |a_1 b_1\rangle \otimes \cdots \otimes |a_n b_n\rangle),
\]
where $a_i = \pm  i$ and $b_i = \pm i$ are eigenvalues for
$\Gamma_{12}$ and $\Gamma_{34}$ respectively in each spinor representation 
that constitutes the tensor product.
The eigenvalue of this state for $G_{12}, G_{34}$ and $G_{5}$ is given by
\[
G_{12} |V\rangle = \sum_i a_i |V\rangle,\  \ 
G_{34} |V\rangle = \sum_i b_i |V\rangle,\  \
G_5 |V\rangle = \sum_i a_i b_i |V\rangle.
\]
For a given state, we can obtain another state with the same $G_5$
eigenvalue by flipping the sign of every $a_i$ and $b_i$.  This new
state has $G_{12}, G_{34}$ eigenvalues which are equal in magnitude
but opposite in sign from those of the original state. Therefore, if
we take a trace over a subspace with a fixed $G_5$ eigenvalue,
$G_{12}$ and $G_{34}$ vanish.

Since the 2-brane charge is proportional to $\Tr G_{ij}$, it follows from the 
above result and rotational symmetry that the 2-brane charge vanishes 
completely everywhere on the 4-sphere.

\subsection{Equations of motion}

By inserting $X_i = (r/n) G_i$ into the Matrix action, 
we obtain the 
the effective Lagrangian
\begin{equation}
L = \frac{N}{2R} \dot{r}^2 - \frac{n R \Omega_4 }{(2\pi)^4 l_{11}^6 } r^4 
+ {\cal O}(1/n).
\label{MatrixLag}
\end{equation}
from which we can derive the equation of motion for the spherical
L5-brane
\begin{equation}
\ddot{r} = -\frac{4n R^2 \Omega_4 }{(2\pi)^4 N l_{11}^6} r^3
\label{eq:eom}
\end{equation}
Factors of $l_{11}$ have been restored so that we can compare
with the predictions of M-theory.

To compare with M-theory, let us recall some recent statements about
the nature of Matrix theory.
Matrix theory can be considered as a definition of the
discrete light-cone quantized (DLCQ) M-theory \cite{Susskind-DLCQ}. One
way to derive the Matrix Hamiltonian from this point of view has been
given by Seiberg \cite{Seiberg-DLCQ}. He considers an M-theory with
Planck length $l_{11}$ compactified on a light-like circle with radius
$R$ and another M-theory ($\tilde{M}$) with Planck length
$\tilde{l}_{11}$ compactified on a spatial circle with radius
$\tilde{R}_s$.  The light-cone Hamiltonian of M-theory is then
identified with the ordinary Hamiltonian of the $\tilde{M}$-theory
minus $N/\tilde{R_s}$ in the limit $\tilde{R}_s, \tilde{l}_{11}
\rightarrow 0$ with $ \tilde{R}_s / \tilde{l}_{11}^2 = R / l_{11}^2$
fixed.  In this limit, $\tilde{M}$-theory becomes Type IIA with
$\alpha' \rightarrow \infty, g_s \rightarrow 0$ and the dynamics is
governed by the Hamiltonian for $N$ slowly moving D0-branes which
translates into the Matrix Hamiltonian.

One can now apply the same principle to the world-volume theory of $n$ 
L5-branes instead of the full M-theory and check whether it leads to the same 
equation of motion. In $\tilde{M}$ theory, $n$ L5-branes with
$p^+ = N/R$ become a bound state of $n$ D4 branes and $N$ D0 branes. 
Such a system is described by the $U(n)$ 
Dirac-Born-Infeld (DBI) action for the D4-branes around a background
gauge field whose spatial field strength has instanton number $N$. 

We assume the following form for the fields
\[
\tilde{X}_i = \tilde{r}(t) n_i(\sigma),\   \ (i=1,\cdots,5)
\]
\[
F_{ab} = \frac{f_{ab}}{\tilde{r}^2},\  \  \;\;\;\;\;
\frac{1}{8\pi^2} \int_{S^4} f \wedge f = N,\  \ \;\;\;\;\;f = *_4 f,
\]
where $\sigma^a$, $a=1,2,3,4$ are the angular coordinates on the unit
4-sphere, $n_i$ is a unit vector on the sphere and $f_{ab}$ is the
dimensionless field strength. $*_4$ denotes the Hodge dual on the
4-sphere.  $\tilde{r}$ is related to the physical radius of M-theory
$r$ by $ \tilde{r}/\tilde{l}_{11} = r/l_{11}$.  With this ansatz, the
DBI Lagrangian becomes
\begin{eqnarray*}
L_{DBI} &=& - \frac{1}{(2\pi)^4 g_s l_s^5} \int d^4\sigma \Tr 
\sqrt{ -\det(G_{ab} + 2\pi l_s^2 F_{ab})}
\cr &=& 
-\frac{1}{(2\pi)^4 g_s l_s^5}\int d^4 \sigma 
\tilde{r}^4 \sqrt{1-\dot{\tilde{r}}^2} 
\Tr \left( 1 + \frac{(2\pi)^2l_s^4}{4\tilde{r}^4} f_{ab} f^{ab} \right).
\end{eqnarray*}
Higher order terms in $f$ from the square root vanish due to 
the self-duality of $f$.

Using the usual $\tilde{M}$-IIA relations 
$\tilde{l}_{11} = g_s^{1/3} l_s$, $\tilde{R}_s = g_s l_s$, 
one can see that only three terms in the Lagrangian survive 
the $\tilde{R}_s$ limit mentioned above,
\[
L_{DBI} = -\frac{N}{\tilde{R}_s} + \frac{N}{2R} \dot{r}^2 
- \frac{nR\Omega_4 r^4}{(2\pi)^4 l_{11}^6}.
\]
The first term is the (divergent) ground state energy we should subtract and
the other two terms are exactly the same as in (\ref{MatrixLag}).

This derivation shows that the equations of motion of the Matrix
theory L5-brane agree with those of the 5-brane of DLCQ M-theory, in
accord with the philosophy of \cite{Seiberg-DLCQ}.  It would be nice
to find a more direct connection with the equations of motion of the
5-brane in 11-dimensional supergravity by starting with a 5-brane
action such as that of \cite{Schwarz-coupling} and deriving the
effective Lagrangian above.  In principle, this should be possible.
The spherical L5-brane with longitudinal momentum corresponds to an
ansatz for the metric tensor $G_{\mu \nu}$ and 2-form field $B_{\mu
\nu}$ on the 5-brane world-volume such that the metric tensor
is that of an embedded 4-sphere with a radius depending only on
$x^+$ while the 3-form field strength $H_{\mu \nu \rho}$ is
independent of $x^-$ and reduces to the field strength of an instanton
on the 4-brane with longitudinal momentum $T^{11}{}_{0} =
H^{11 \mu \nu} H_{\mu \nu 0}$.  Such a derivation is left for future
work.

\subsection{Long-range interactions}

The long-range gravitational field of the spherical L5-brane can be
calculated by considering the interaction of the brane with a
graviton.  The leading long-distance potential between the L5-brane
and a graviton can be determined using the standard one-loop Matrix
theory calculation.  The general form of the interaction potential
between any two Matrix theory objects was described in
\cite{ChepelevTseytlin1,Dan-Wati} and is given by\footnote{In this and
the following subsection, we follow the units of
\cite{Dan-Wati}, in which $2\pi l_{11}^3 = R$.}
\[
V_{{\rm matrix}} = -\frac{5}{128b^7}  W
\]
where $b$ is the separation distance between the objects and where the
``gravitational coupling'' $W$ is given by
\begin{equation}
W = \Tr\left[
   8 F^\mu{}_\nu F^\nu{}_\lambda F^\lambda{}_\sigma F^\sigma{}_\mu
+ 16 F_{\mu \nu} F^{\mu\lambda} F^{\nu\sigma} F_{\lambda \sigma}
-  4 F_{\mu\nu} F^{\mu \nu} F_{\lambda \sigma} F^{\lambda \sigma}
-  2 F_{\mu\nu} F_{\lambda \sigma} F^{\mu\nu} F^{\lambda \sigma}
\right ].
\label{eq:W}
\end{equation}
Taking the probe graviton to be stationary in the transverse
directions, the field strength components are given by
$F_{0i}=-F_{i0}=\partial_{t}X_{i}$ and $F_{ij}=i\left [ X_{i},X_{j}
\right]$.

The interaction potential can be computed exactly using the identities
in Appendix A.  
The gravitational coupling for the L5-brane-graviton system is given by
\begin{eqnarray}
W&= &24N\left [ 0 \cdot
\frac{r^{8}}{n^{8}}+16 \cdot
\frac{r^{4}\dot{r}^{2}}{n^{6}}+\frac{\dot{r}^{4}}{n^4}
	\right ]c^2  -64 N \left [48 \cdot
\frac{r^{8}}{n^{8}}-8 \cdot
\frac{r^{4}\dot{r}^{2}}{n^{6}}+\frac{\dot{r}^{4}}{n^4}
	\right ]c\label{eq:potential}\\
& = & 96 N \left[ \frac{\dot{r}^4}{4}  + 4
\frac{r^4 \dot{r}^2}{n^2}  \right] \cdot (1 +{\cal O} (1/n)). \nonumber
\end{eqnarray}

The terms which appear at leading order in $n$ can also be computed
directly.  To leading order in $n$, the last two terms in (\ref{eq:W})
are proportional to $(F_{\mu \nu }^{2})^{2}$, and will therefore be
proportional to $c^2$ where $c$ is the Casimir computed in
(\ref{eq:Schur}).  For large $n$, the first two terms in $W$ will be
dominated by the $n(n-1)(n-2)(n-3)$ terms of the form $\Gamma \otimes
\Gamma \otimes \Gamma \otimes \Gamma \otimes \identity \otimes \dots
\otimes \identity +sym.$, that is, those terms with each $\Gamma$
matrix in a different tensor block.  These terms can be easily
computed and shown to sum to the leading order term above.

This result can be compared to the interaction potential computed in
supergravity.  The supergravity potential describing the interaction
due to single graviton exchange (with no longitudinal momentum
transfer) between a massive object and a graviton with $p^+ = 1/R,
p^-= 0$ in light-front coordinates is given by \cite{Dan-Wati}
\[
V_{\rm gravity} = -\frac{15}{4}  \frac{R}{b^7} \;T^{--}
\]
where $T^{--}$ is the $--$ component of the stress-energy tensor of
the massive object
integrated over $dx^- d^{9}x^T$.  The spherically symmetric membrane
discussed in \cite{Dan-Wati} has momentum $p^-$ uniformly distributed
over the membrane world-volume and can be treated as a pointlike
object with $T^{--}= p^-p^-/p^+ = E^2 R/N$ where $E$ is the
light-front (Matrix theory) energy of the membrane.  The story is
similar for the L5-brane; however, because the L5-brane is extended in
the longitudinal direction there is an additional contribution to
the $--$ component of the stress-energy tensor from the brane tension,
so that
\[
T^{--} = \frac{R}{N}  E^2 - \beta r^8.
\]
The coefficient $\beta$ can be fixed by noting that a configuration
containing a relatively stationary 4-brane and 0-brane is
supersymmetric \cite{Polchinski-TASI}.  Thus, the term in the
stress-energy tensor proportional to L5-brane charge squared must
vanish.

Calculating the Matrix theory energy of the spherical L5-brane
\begin{eqnarray*}
E&=&\frac{1}{R} \Tr \left(\frac{1}{2} \dot{X}_{i} \dot{X}_{i} -\frac{1}{4}
\left[ X_{i},X_{j} \right]^{2} \right) \\ 
 &=&\frac{N}{R} \left(\frac{\dot{r}^{2}}{2n^2}+4\frac{r^{4}}{n^{4}}
\right) c = \frac{N}{R} 
\left(\frac{\dot{r}^{2}}{2}+4\frac{r^{4}}{n^{2}}
\right) \cdot (1 +{\cal O} (1/n))
\end{eqnarray*}
we see that the Matrix theory and supergravity potentials are in exact
agreement at leading order.

We have seen that the term in the Matrix theory potential proportional
to $r^8$ vanishes at leading order in $n$, as we would expect from the
fact that there is no static force between a 4-brane and a 0-brane.
In the context of the DLCQ form of the Matrix theory conjecture, it is
natural to expect that this term should vanish to all orders in $n$.
However, this is not the case.  The subleading term in
(\ref{eq:potential}) proportional to $c$ contains a term proportional
to $r^8$.  The discrepancy this suggests between Matrix theory and
DLCQ supergravity is analogous to the finite $N$ discrepancy noted in
\cite{Dan-Wati} for the membrane-graviton potential, and gives further
evidence that there may be a distinction between the DLCQ of
11-dimensional supergravity and the low-energy DLCQ of M-theory
\cite{Seiberg-DLCQ,Hellerman-Polchinski}.

\subsection{Fluctuations}
\label{sec:fluctuations}

We have successfully constructed longitudinal 5-branes whose geometry
in Matrix theory corresponds to a completely symmetric 4-sphere.  It
is natural to want to extend this type of description to an arbitrary
geometry with the topology of a 4-sphere.  In the case of the Matrix
membrane, a description of a general membrane geometry with spherical
topology was given by de Wit, Hoppe and Nicolai \cite{dhn}.  Their
general construction was based on the observation that an embedding of
the spherical membrane world-volume can be approximated arbitrarily
well by a polynomial in the coordinates $x_1, x_2, x_3$ on the
world-volume of the sphere.  They defined the natural map from
polynomials in these coordinates to symmetric polynomials in the
$SU(2)$ generators $J_i$, and showed that the Matrix equations of
motion correspond to the membrane equations of motion in the large $N$
limit.  This equivalence follows from the fact that the matrix
commutator $[\cdot,
\cdot]$ has the same action on symmetrized polynomials in the $SU(2)$
generators as the Poisson bracket $\{\cdot, \cdot\}$ does on the
corresponding polynomials in the coordinates $x_i$.

We would like to find a corresponding picture for the longitudinal
5-brane, which would allow us to describe the classical dynamics of an
arbitrary L5-brane geometry.  However, it is not possible to simply
generalize the membrane results by considering symmetrized polynomials
in the matrices $G_i$.  We will now discuss briefly the problems
which arise in attempting to do this.  A straightforward
generalization of the membrane description of de Wit, Hoppe and
Nicolai might indicate that we could include a fluctuation in the
membrane configuration by simply adding higher order symmetric
polynomials in the $G$'s to the Matrix configuration
(\ref{eq:sphere}).  In order for this proposal to work, it is
necessary that the accelerations $\ddot{X^i}$ calculated through the
equations of motion 
\[
\ddot{X}_i = -[[X_i, X_j], X_j]
\]
should themselves be expressable in
terms of symmetrized polynomials in the $G$'s.  Indeed, this is
generally not possible.  We will now demonstrate this in a simple
case.  Let us consider  an L5-brane which is embedded at $t = 0$ 
according to a nonlinear function of the matrices $G$
\begin{eqnarray*}
X^1 & = &  \frac{r^2}{2n^2}
 \left(G_1 G_2+G_2 G_1 \right)\\
X^2 & = &  \frac{r}{n} G_3.
\end{eqnarray*}
The acceleration of $X^1$ can be found from the equations of motion
\[
\ddot{X}^1 = -[[X^1, X^2], X^2].
\]
Let us analyze the structure of these expressions for arbitrary $n$.
The symmetrized product in $X^1$ gives rise to a matrix containing
terms of the form $\Gamma_1 \otimes \Gamma_2 + \Gamma_2 \otimes
\Gamma_1$ acting on the various pairs of spaces in the tensor
product.  The repeated commutator with $X^2$ gives two types of terms:
the first is simply proportional to $X^1$; the second is of the form
\begin{equation}
\Gamma_{13} \otimes \Gamma_{23} + \Gamma_{23} \otimes \Gamma_{13}
\label{eq:problem}
\end{equation}
acting on all pairs of spaces in the tensor product space.  In order
to find a polynomial expression for $\ddot{X}^1$ it would be necessary
to express (\ref{eq:problem}) as a polynomial in the $G's$.  However,
this does not seem to be possible.  Any polynomial containing terms of
higher than quadratic order would have a matrix tensor expression in
which there was a nontrivial action on more than two of the blocks in
the tensor product space.  And an explicit check of all linear and
quadratic polynomials indicates that (\ref{eq:problem}) cannot be
expressed as a linear combination of such terms.

Thus, it seems that the Matrix equations of motion do not close on the
class of symmetric polynomials in the generating matrices $G_i$.  This
makes it difficult to imagine how one might incorporate a generally
fluctuating L5-brane into the framework we have described here.  A
further complication arises from the fact that our spherical membranes
have a winding number $n$ which is related to the number of 0-branes
$N$ used in the construction.  Since the solutions we have given
describe multiple spherical membranes, some of the degrees of
freedom of the system should correspond to modes where some of the
spheres become smaller than others.  In the large radius limit, these
degrees of freedom should correspond to zero-modes of the system.
However, these degrees of freedom do not seem to appear naturally in
the spherical membranes we have constructed.  For example, the double
sphere with $n = 2$ contains $N = 10$ 0-branes.  It should be possible
to split the sphere into two independent membrane spheres of radii $r
\pm \epsilon$.  However, we only know of a single sphere construction
for $N = 4$, so that somehow two 0-branes would have to be annihilated
in the course of this decomposition.

\section{Conclusions}

We have given an explicit construction of longitudinal 5-branes with
transverse spherical geometry in Matrix theory.  These spherical
branes have many of the physical properties we expect from M-theory.
However, there is an unusual relationship between the number of
0-branes needed to construct the 5-brane and the winding number of the
5-brane configuration.  Unlike the spherical matrix membrane, the
4-sphere brane solutions cannot be constructed for arbitrary $N$.
Furthermore, it seems to be difficult to understand the fluctuation
spectra of the spherical 5-branes in terms of Matrix theory variables.
The difficulties we have encountered here do not prove conclusively
that there is no Matrix theory description of a fluctuating
longitudinal 5-brane.  However, they indicate that it will not be
straightforward to find a description of the longitudinal 5-brane with
arbitrary geometry.  It may be that the best way to find such a
description would involve a direct quantization in light-front
coordinates of a 5-brane action.  This problem is left as a challenge
for further research.

It would also be interesting to try to find a construction analogous to
that given here for a transverse Matrix 5-brane.  Although the charge
for such an object does not seem to appear in the Matrix theory
supersymmetry algebra \cite{bss}, an implicit description of a flat
transverse 5-brane was given in \cite{grt} using super Yang-Mills
S-duality in $D = 4$.  If a set of matrices could indeed be found
describing 0-branes localized on a 5-sphere, it might give some
interesting insights into the nature of the transverse 5-brane in
Matrix theory.

\section*{Acknowledgements}

We would like to thank Dan Kabat, Morten Krogh, Shiraz Minwalla, Jaemo Park,
Larus Thorlacius and Dan Waldram for helpful discussions.
The work of SL is supported in part by the Department of Energy 
(DOE) under contract DE-FG02-91ER40671.  
The work of WT is supported in part by the National Science Foundation
(NSF) under contract PHY96-00258.

\newpage

\section*{Appendix A. Some matrix identities}

Let us begin with the definitions of the matrices
\begin{eqnarray*}
G_i &=& \left(
\Gamma_i \otimes \identity \otimes \cdots \otimes \identity +
\identity \otimes \Gamma_i \otimes \identity \otimes 
\cdots \otimes \identity + \cdots +
\identity\otimes \cdots \otimes\identity \otimes\Gamma_i \right)_{\rm Sym},
\cr
G_{ij} &=& \half [G_i,G_j]  
\end{eqnarray*}
The dimension and the ``Casimir'' of the $n$-fold totally symmetric 
representation are given by
\[
N=\frac{(n+1)(n+2)(n+3)}{6},\  \ c = n (n+4).
\]
The Casimir, $c$, appears in the following identities:
\begin{equation}
G_i G_i = c \identity_N, \;\;\;\;\;\  \ G_{ij} G_{ji} = 4 c \identity_N.
\label{eq:Schur}\end{equation}
These follow from the basic identities
\begin{eqnarray*}
(\Gamma_i \otimes \Gamma_i)_{\rm Sym} = (\identity \otimes
\identity)_{\rm Sym},
\;\;\;\;\;
(\Gamma_{ij} \otimes \Gamma_{ji})_{\rm Sym} = 
4 (\identity \otimes \identity)_{\rm Sym}.
\label{Schure}
\end{eqnarray*}
The fact that these matrices are proportional to the identity matrix
is a result of Schur's lemma. The proportionality constant can be
determined by multiplying both sides by any vector in the symmetric
representation in a specific basis.

The commutators of $G_i$'s and $G_{jk}$'s are easily obtained from those of 
gamma matrices:
\begin{eqnarray*}
[G_{ij},G_k] &=& 2(\delta_{jk} G_i - \delta_{ik} G_j),
\cr
[G_{ij}, G_{kl}] &=& 
2(\delta_{jk}G_{il} + \delta_{il}G_{jk} 
- \delta_{ik}G_{jl} - \delta_{jl}G_{ik}).
\label{commutator}
\end{eqnarray*}
There is a very useful identity between the anti-commutators: 
\[
\{ G_{ij}, G_{jk} \} + \{ G_i, G_k \} = 2c\delta_{ik} \identity_N.
\]
To prove this, we need to use the Fierz identity among the gamma matrices of
${\rm Spin}(5)$ in addition to Schur's lemma. We omit the details.

Combining the above definitions and identities, it is straighforward to derive
the following trace formulas that can be used in the calculation of the 
gravitational interaction.
\begin{eqnarray*}
\Tr(G_i G_j G_i G_j) = N(c^2 - 8c)  &\hspace{0.5in} &
\Tr(G_i G_j G_{ik} G_{kj}) = -8Nc
\\
\Tr(G_i G_{jk} G_i G_{jk}) =N(-4 c^2 + 16 c)  & &
\Tr(G_{ij} G_{kl} G_{ij} G_{kl}) =N (16c^2 - 96 c)
\\
\Tr(G_{ij} G_{jk} G_{kl} G_{li}) = N(4c^2 + 32c) & &
\Tr(G_{ij} G_{jk} G_{il} G_{lk}) = N(4c^2 - 40c)
\end{eqnarray*}
\[
\Tr(G_i G_j G_{jk} G_{ki}) = \Tr(G_i G_{ik} G_j G_{jk}) 
= \Tr(G_i G_{jk} G_j G_{ik}) =16Nc
\]
\section*{Appendix B. Normalization conventions}

The 11 dimensional Planck length $l_{11}$ is defined by
\[
16 \pi G_{11} = (2\pi)^8 l_{11}^9,
\]
where $G_{11}$ is the 11-dimensional Newton's constant. In this convention,
the tensions of the membrane and the 5-brane take the simple forms
\[
T_{M2} = \frac{1}{(2\pi)^2 l_{11}^3},\ \; \;\;\;\;
T_{M5} = \frac{1}{(2\pi)^5 l_{11}^6}.
\]
The bosonic part of the Matrix theory Hamiltonian and Lagrangian read
\[
H = R\;\Tr\left\{ \half \Pi_i^2 
- \frac{1}{4 (2\pi)^2 l_{11}^6} [X_i,X_j]^2\right\},\;\;\;\;\;
L = \Tr\left\{ \frac{1}{2R}\dot{X}_i^2 
+ \frac{R}{4 (2\pi)^2 l_{11}^6} [X_i,X_j]^2 \right\}.
\]
We choose a convention for the string coupling $g_s$ of 
the Type IIA theory in such a way that the following relations are exact
\[
l_{11} = g_s^{1/3} l_s,\  \ R_{11} = g_s l_s,
\]
where $l_s = \sqrt{\alpha'}$ is the string length and $R_{11}$ is the 
compactification radius. With this choice, the
tension of the D-$p$-branes become
\[
T_{Dp} = \frac{1}{(2\pi)^p g_s l_s^{p+1}}.
\]
The low energy dynamics of the D-branes are governed by the DBI action:
\[
S_{DBI} = - T_{Dp} \int d^{p+1}\sigma \Tr
\sqrt{ -\det(G_{ab} + B_{ab} + 2\pi l_s^2 F_{ab})}.
\]

\newpage

\bibliographystyle{plain}


\end{document}